\documentclass[aps,prd,floatfix,groupedaddress,nofootinbib,superscriptaddress,twocolumn]{revtex4-2}

\usepackage{amsmath,amssymb,bm,color,float,graphicx,mathrsfs}
\usepackage[hidelinks]{hyperref}
\hypersetup{colorlinks=true,linkcolor=blue,citecolor=blue,urlcolor=blue}
\usepackage{Macro}

\def\chie{\chi_{\rm e}}

\begin{document}

\title{Cosmic birefringence tomography with polarized Sunyaev Zel'dovich effect}

\author{Toshiya Namikawa}
\affiliation{Center for Data-Driven Discovery, Kavli IPMU (WPI), UTIAS, The University of Tokyo, Kashiwa, 277-8583, Japan}

\author{Ippei Obata}
\affiliation{Kavli IPMU (WPI), UTIAS, The University of Tokyo, Kashiwa, 277-8583, Japan}

\date{\today}

\begin{abstract}
We consider the polarized Sunyaev-Zel'dovich (pSZ) effect for a tomographic probe of cosmic birefringence, including all relevant terms of the pSZ effect in the cosmic microwave background (CMB) observables, some of which were ignored in the previous works. The pSZ effect produces late-time polarization signals from the scattering of the local temperature quadrupole seen by an electron. We forecast the expected constraints on cosmic birefringence at the late time of the universe with the pSZ effect. We find that the birefringence angles at $2\alt z\alt 5$ are constrained at a subdegree level by the cross-correlations between CMB $E$- and $B$-modes or between CMB $B$-modes and remote quadrupole $E$-modes using data from LiteBIRD, CMB-S4, and LSST. In particular, the cross-correlation between large-scale CMB $B$-modes and remote-quadrupole $E$-modes has a much smaller bias from the Galactic foregrounds and is useful to cross-check the results from the $EB$ power spectrum. 
\end{abstract} 



\maketitle

\section{Introduction} \label{sec:intro}

Cosmic birefringence --- a rotation of the linear polarization plane of the cosmic microwave background (CMB) as they travel through space\footnote{The nomenclature of this rotation effect is discussed in \cite{Ni:2007:cpr}.} --- is now a key observable to search for parity-violating physics in cosmology \cite{Komatsu:2022:review}. 
Recent measurements of the cross-correlation between the even-parity $E$-modes and odd-parity $B$-modes in the polarization map suggest a tantalizing hint of cosmic birefringence  \cite{Minami:2020:biref,Diego-Palazuelos:2022,Eskilt:2022:biref-freq,Eskilt:2022:biref-const,Eskilt:2023:cosmoglobe}.
Cosmic birefringence can be induced by a pseudoscalar field, such as axionlike particles (ALPs), coupled with electromagnetic fields via the so-called Chern-Simons term, $\mC{L}\supset -g_{\phi\gamma}\phi F^{\mu\nu}\tilde{F}_{\mu\nu}/4$, where $g_{\phi\gamma}$ is the ALP-photon coupling constant, $\phi$ is an ALP field, $F^{\mu\nu}$ is the electromagnetic field tensor, and $\tilde{F}_{\mu\nu}$ is its dual. 
Cosmic birefringence can be caused by the ALP field of dark energy~\cite{Carroll:1998:DE,Panda:2010,Fujita:2020aqt,Fujita:2020ecn,Choi:2021aze,Obata:2021,Gasparotto:2022uqo,Galaverni:2023}, early dark energy \cite{Fujita:2020ecn,Murai:2022:EDE,Eskilt:2023:EDE}, dark matter~\cite{Finelli:2009,Liu:2016dcg,Fedderke:2019:biref}, and by topological defects \cite{Takahashi:2020tqv,Kitajima:2022jzz,Jain:2022jrp,Gonzalez:2022mcx}, as well as by possible signatures of quantum gravity~\cite{Myers:2003fd,Arvanitaki:2009fg}. 
Upcoming CMB experiments, including the BICEP \cite{Cornelison:2022:BICEP3,BICEPArray}, Simons Array \cite{SimonsArray}, Simons Observatory \cite{SimonsObservatory}, CMB-S4 \cite{CMBS4}, and LiteBIRD \cite{LiteBIRD}, with which the polarization noise will be reduced significantly, are expected to improve cosmic birefringence measurements. 

Multiple studies have shown that the shape of the $EB$ power spectrum depends on the dynamics of the ALP fields during reionization and recombination \cite{Finelli:2009,Sigl:2018:biref-sup,Sherwin:2021:biref,Nakatsuka:2022}, including early dark energy \cite{Murai:2022:EDE,Eskilt:2023:EDE,Yin:2023:biref}, dark energy \cite{Liu:2006,Lee:2013:const}, and other phenomenological models \cite{Gubitosi:2014:biref-time}. Hence, measuring the spectral shape of the power spectrum will provide tomographic information on such scenarios. This method, the {\it cosmic birefringence tomography}, can avoid the degeneracies with the instrumental miscalibration angle \cite{QUaD:2008ado,Miller:2009pt,Komatsu:2010:WMAP7,B1rot,P16:rot} and half-wave plate nonidealities \cite{Monelli:2022:HWP}. 

This paper considers a new tomographic source --- the polarized Sunyaev-Zel'dovich (pSZ) effect, which generates linear polarization through Thomson scattering of CMB temperature quadrupole by free electrons in clusters or intergalactic space in the late time of the universe \cite{Kamionkowski:1997:pSZ,Portsmouth:2004:pSZ,Seto:2005,Hall:2014,Louis:2017:pSZ}. Measuring the polarization signal from the pSZ effect provides information on cosmic birefringence in the late-time universe. The pSZ-induced polarization signal is usually expressed by the remote quadrupole fields, which are decomposed into $E$- and $B$-modes, $q^E$ and $q^B$ (hereafter, remote quadrupole $E$- and $B$-modes). Reference \cite{Alizadeh:2012:pSZ-est} provides an estimator to reconstruct $q^E$ and $q^B$ by cross-correlating observed CMB $E$- or $B$-modes with large-scale structure tracers, such as galaxy number density fluctuations. Future CMB experiments, such as CMB-S4 \cite{CMBS4} and CMB-HD \cite{CMBHD}, with future galaxy surveys, such as the Vera Rubin Observatory Legacy Survey of Space and Time (LSST) \cite{LSST}, would be able to detect the remote quadrupole \cite{Deutsch:2017:pSZ-rec}. Multiple studies have discussed applications of the remote quadrupole for cosmology, including the large-scale CMB anomalies \cite{Cayuso:2019}, the integrated Sachs-Wolfe effect \cite{Cooray:2002:ISW,Bunn:2006:ISW}, CMB optical depth \cite{Meyers:2017:pSZ}, and inflationary gravitational waves \cite{Deutsch:2018:IGW}. 

Recently, Hotinli {\it et al.} \cite{Hotinli:2022:pkSZ} and Lee {\it et al.} \cite{Lee:2022:pSZ-biref} have considered the birefringence effect on $q^E$ and $q^B$ to constrain cosmic birefringence in the late-time universe. 
The remote quadrupole is tiny, however, and the expected constraints on the birefringence angle from even next-generation CMB experiments and galaxy surveys are at the level of degrees to $10$ degrees. 
In this paper, we further consider the pSZ-induced polarization in the observed CMB $E$- and $B$-modes and explore how the constraints on the birefringence angle improve by including these new contributions in conjunction with $q^E$ and $q^B$. 

This paper is organized as follows. Section \ref{Sec:pSZ} reviews the pSZ effect and formulates the pSZ effect in the presence of cosmic birefringence. Section \ref{Sec:forecast} shows the expected constraint on cosmic birefringence by combining large-scale CMB polarization and remote quadrupole. Section \ref{Sec:Conclusion} is devoted to a conclusion. 

\begin{table}[t]
 \centering
 \begin{tabular}{l|ccc}
 Experiment & $\sigma_{\rm P}$ & $\theta_{\rm FWHM}$ & $A_{\rm lens}$ \\
 & [$\mu$K-arcmin] & [arcmin] & \\ \hline
 LiteBIRD & 2 & 30 & \\ \hline
 S4 & 1 & 1.4 & 0.2 \\
 HD & 0.4 & 0.2 & 0.1 \\
 \end{tabular}
 \caption{Setup for a LiteBIRD-like (LiteBIRD), CMB-S4-like (S4), and CMB-HD-like (HD) experiments. $\sigma_{\rm P}$ is the map noise level in $\mu$K-arcmin, $\theta_{\rm FWHM}$ is the FWHM of the Gaussian beam in arcmin, and $A_{\rm lens}$ is the fraction of the residual lensing $B$-mode spectrum after delensing with that experiment. LiteBIRD measures large-scale CMB polarization, and S4/HD reconstructs the remote quadrupole. For LiteBIRD, we assume delensing with a reconstructed lensing map from ground-based experiments and choose $A_{\rm lens}=0.2$ with S4 and $0.1$ with HD.}
 \label{tab:cmbexp}
\end{table}

Throughout this paper, we define the spherical harmonic decomposition of a spin-$0$ quantity, $x$, as
\al{
    x_{\l m} = \Int{2}{\hatn}{} Y^*_{\l m}(\hatn) x(\hatn)
    \,, 
}
where $Y_{\l m}$ is the spherical harmonics. We also define the $E$- and $B$- modes from the Stokes $Q$ and $U$ parameters \cite{Zaldarriaga:1996xe,Kamionkowski:1996:eb}:
\al{
    E_{\l m} \pm\iu B_{\l m} = - \Int{2}{\hatn}{} (Y_{\l m}^{\pm2}(\hatn))^* P^\pm(\hatn)
    \,, \label{Eq:EB-def}
}
where $P^\pm=Q\pm\iu U$ and $Y_{\l m}^{\pm2}$ is the spin-2 spherical harmonics. 
We assume the flat $\Lambda$CDM cosmology obtained from Planck \cite{P18:main}. 
The experimental configuration for CMB used in this paper is summarized in Table \ref{tab:cmbexp}.

\section{Polarized SZ effect} \label{Sec:pSZ}

In this section, we briefly review the pSZ effect by following \cite{Deutsch:2017:pSZ-cl} and discuss the cosmic birefringence effect on the polarization signals generated by the pSZ effect. 

\subsection{Remote quadrupole}

In CMB observations, we measure the Stokes $Q$ and $U$ maps along the line-of-sight direction, $\hatn$. The Stokes $Q$ and $U$ map are given by \cite{Deutsch:2017:pSZ-cl} 
\al{
    P^\pm(\hatn) = -\INT{}{\chi}{}{0}{\chi_*} g_{\rm vis}(\chi) \frac{\sqrt{6}}{10}q^\pm(\chi,\hatn) 
    \,. \label{Eq:Pol}
}
Here, $\chi_*$ is the comoving distance from an observer to the last scattering surface of CMB, and we define the visibility function as
\al{
    g_{\rm vis}(\chi) = \D{\tau}{\chi}\E^{-\tau(\chi)} = \sigma_{\rm T}a(\chi)n_{\rm e}(\chi) \E^{-\tau(\chi)}
    \,, 
}
where $\sigma_{\rm T}$ is the cross section of the Thomson scattering, $a$ is the scale factor, and $n_{\rm e}$ is the electron number density. The CMB optical depth, $\tau$, is defined as
\al{
    \tau(\chi) = \INT{}{\chi'}{}{0}{\chi} \sigma_{\rm T}a(\chi')n_{\rm e}(\chi')
    \,.
}
The remote quadrupole fields, $q^\pm(\chi,\hatn)$, are decomposed into $E$- and $B$-modes, $q^E_{\l m}(\chi)$ and $q^B_{\l m}(\chi)$, using Eq.~\eqref{Eq:EB-def}. If we consider only the linear density perturbations, $q^B_{\l m}(\chi)$ vanishes \cite{Philcox:2022:psz}. On the other hand, $q^E_{\l m}(\chi)$ is related to the primordial gravitational potential as \cite{Deutsch:2017:pSZ-cl}
\al{
    q^E_{\l m}(\chi) = 4\pi \Int{3}{\bm{k}}{(2\pi)^3} \Delta^{q^E}_\l(k,\chi) \Psi_i(\bm{k})Y^*_{\l m}(\hk) \,,
}
with
\al{
    \Delta_\l^{q^E}(k,\chi) &= 5\iu^\l\sqrt{\frac{3}{8}\frac{(\l+2)!}{(\l-2)!}}\frac{j_\l(k\chi)}{(k\chi)^2}T(k)
    \notag \\
    &\qquad \times \sum_{{\rm X}={\rm SW},{\rm ISW},{\rm Doppler}}\mC{G}_{\rm X}(k,\chi)
    \,. 
}
Here, $\Psi_i$ is the primordial gravitational potential, $T(k)$ is the transfer function, and
\al{
    \mC{G}_{\rm SW} &= -\left(2D_{\Psi}(\chi_*)-\frac{3}{2}\right)j_2(k(\chi_*-\chi)) 
    \,, \\
    \mC{G}_{\rm ISW} &= -2\INT{}{a}{}{a_*}{a_{\rm e}}\D{D_{\Psi}}{a}j_2(k(\chi-\chi))
    \,, \\
    \mC{G}_{\rm Doppler} &= \frac{k}{5}D_v(\chi_*)[3j_3(k(\chi_*-\chi))-2j_1(k(\chi_*-\chi))]
    \,. 
}
$D_\Psi$ is the growth function of the gravitational potential computed with the analytic formula of \cite{Erickcek:2008}. $D_v$ is the velocity growth factor and is given by
\al{
    D_v(\chi) \equiv \frac{2a^2H(\chi)}{H_0^2\Omega_{\rm m}}\frac{y}{4+3y}\left(D_\Psi+\D{D_\Psi}{\ln a}\right)
    \,, 
}
where $H_0$ is the expansion rate at present, $\Omega_{\rm m}$ is the fractional energy density of the matter component at present, $H(\chi)=H_0\sqrt{\Omega_{\rm m}a^{-3}+1-\Omega_{\rm m}}$, and $y=a/a_{\rm eq}$ with $a_{\rm eq}$ being the radiation-matter equality time. We set $a=1$ at the present epoch.
The angular power spectrum of the remote-quadrupole $E$-modes is given by
\al{
    C_{\l}^{q^Eq^E}(\chi,\chi') = 4\pi\Int{}{\ln k}{}\mC{P}_\Psi(k) \Delta_\l^q(k,\chi)\Delta_\l^q(k,\chi')
    \,, 
}
where $\mC{P}_\Psi(k)$ is the dimensionless power spectrum of $\Psi_i$.

In CMB observations, the observed polarization contains contributions of polarization generated at the late-time universe by the pSZ effect. From Eq.~\eqref{Eq:Pol}, the $E$-mode contribution is written in terms of $q^E$ as 
\al{
    E_{\l m} = -\frac{\sqrt{6}}{10}\INT{}{\chi}{}{0}{\chi_*}g_{\rm vis}(\chi)q^E_{\l m}(\chi)
}
The $E$-mode angular power spectrum is then given by
\al{
    C_\l^{EE} = \frac{6}{100}\INT{}{\chi}{}{0}{\chi_*}\INT{}{\chi'}{}{0}{\chi_*}g_{\rm vis}(\chi)g_{\rm vis}(\chi')C_{\l}^{q^Eq^E}(\chi,\chi')
    \,. 
}

\subsection{Reconstruction of the remote quadrupole}

Next, we review the reconstruction of the remote quadrupole by combining CMB experiments and galaxy surveys, following \cite{Deutsch:2017:pSZ-rec}. The key idea of the reconstruction is that the fluctuations of electron number density modulate the remote quadrupole fields, and this modulation traces the underlying matter density fluctuations. Thus, the remote quadrupole fields are reconstructed from a correlation between this modulation and a large-scale structure tracer.

If the electron number density has fluctuations, the CMB polarization from the pSZ is distorted as
\al{
    \delta P^\pm (\hatn) = -\INT{}{\chi}{}{0}{\chi_*} \bar{g}_{\rm vis}(\chi)\delta_{\rm e}(\chi,\hatn) \frac{\sqrt{6}}{10}q^\pm(\chi,\hatn)
    \,. 
}
Here, $\delta_{\rm e}$ is the fluctuations of the electron number density. We ignore the fluctuations of the screening, $\E^{-\tau(\chi)}$, which are much smaller than the fluctuations of $n_{\rm e}$ well after the reionization \cite{Deutsch:2017:pSZ-cl}. For a given interval of the comoving distance corresponding to the redshift bin in practice, we define the average components of the remote-quadrupole $E$-modes and optical depth in each bin as follows:
\al{
    q^{\pm,i}(\hatn) &= \frac{1}{\Delta\chi_i}\INT{}{\chi}{}{\chi_{i-1}}{\chi_i}q^\pm(\chi,\hatn) 
    \,, \\
    \delta\tau^i(\hatn) &= \INT{}{\chi}{}{\chi_{i-1}}{\chi_i}\bar{g}_{\rm vis}(\chi)\delta_{\rm e}(\chi,\hatn)
    \,,
}
where $\Delta\chi_i$ is the bin width in the comoving distance at the $i$th bin. The distortion to the observed CMB polarization is then given as \cite{Deutsch:2017:pSZ-rec} 
\al{
    \delta P^\pm (\hatn) \simeq -\sum_i \frac{\sqrt{6}}{10}\delta\tau^i(\hatn)q^{\pm,i}(\hatn)
    \,. 
}
Defining the averaged remote-quadrupole $E$- and $B$-modes, $q^{E,i}_{\l m}$ and $q^{B,i}_{\l m}$, using $q^{\pm,i}(\hatn)$, the observed CMB $E$- and $B$- modes involving $\delta_{\rm e}$ are then given by \cite{Deutsch:2017:pSZ-rec}
\al{
    (\delta X_{\l m})^* &= -\frac{\sqrt{6}}{10}\sum_i\sum_{\l_1m_1\l_2m_2}\Wjm{\l}{\l_1}{\l_2}{m}{m_1}{m_2}\gamma_{\l\l_1\l_2}
    \notag \\
    &\times \Wjm{\l}{\l_1}{\l_2}{2}{-2}{0}\sum_{Y=E,B} w^{q^X Y}_{\l\l_1\l_2} q^{Y,i}_{\l_1m_1} \delta\tau^i_{\l_2m_2} 
    \,. \label{Eq:mode-mixing}
}
Here, the large parentheses denote the Wigner-$3j$ symbol, and we define
\al{
    \gamma_{\l \l_1\l_2} &= \sqrt{\frac{(2\l+1)(2\l_1+1)(2\l_2+1)}{4\pi}} 
    \,, \\
    w^{q^E E}_{\l\l_1\l_2} &= \wp^+_{\l\l_1\l_2}
    \,, \\
    w^{q^E B}_{\l\l_1\l_2} &= \iu \wp^-_{\l\l_1\l_2}
    \,, \\
    w^{q^B E}_{\l\l_1\l_2} &= -\iu \wp^-_{\l\l_1\l_2}
    \,, \\
    w^{q^B B}_{\l\l_1\l_2} &= \wp^+_{\l\l_1\l_2}
    \,, 
}
where $\wp^\pm_{\l\l_1\l_2}=[1\pm(-1)^{\l+\l_1+\l_2}]/2$. 

From Eq.~\eqref{Eq:mode-mixing}, we can construct estimators for the remote-quadrupole $E$- and $B$-modes, $q^E$ and $q^B$, from measurements of the CMB $E$- or $B$-modes and a tracer of the density perturbations which correlate with $\delta\tau^i$. The estimator is described as (e.g., \cite{Alizadeh:2012:pSZ-est,Deutsch:2017:pSZ-rec}) \footnote{The minimum variance estimator of \cite{Alizadeh:2012:pSZ-est,Deutsch:2017:pSZ-rec} can be expressed as a linear combination of even and odd parity contributions. The even and odd parity terms are not correlated, and the minimum-variance estimator is given by the inverse-variance sum of these two estimators.}
\al{
    (\hat{q}^{X,i}_{\l m})^* &= N^{q^{X,i}}_\l \sum_{Y=E,B}\sum_{\l_1m_1\l_2m_2}
    \notag \\
    &\times \Wjm{\l}{\l_1}{\l_2}{m}{m_1}{m_2}f^{q^{X,i}Y}_{\l\l_1\l_2}\frac{Y_{\l_1m_1}}{\hat{C}_{\l_1}^{YY}}\frac{(x^i_{\l_2m_2})^*}{\hat{C}^{x^ix^i}_{\l_2}}
    \,, 
}
where $\hat{C}^{YY}_\l$ is the observed power spectrum of $Y$ and $\hat{C}^{x^ix^i}_\l$ is that of the mass tracer at the $i$th bin, $x^i$. The weight function is defined as (e.g., \cite{Philcox:2022:psz})
\al{
    f^{q^{X,i}Y}_{\l\l_1\l_2} &= -\frac{\sqrt{6}}{10}\gamma_{\l\l_1\l_2}\Wjm{\l}{\l_1}{\l_2}{2}{-2}{0}C_{\l_2}^{\delta\tau^i x^i}w^{q^X Y}_{\l\l_1\l_2}
    \,. 
}
The estimator normalization is defined as
\al{
    \frac{1}{N^{q^{X,i}}_{\l}} &= \frac{1}{2\l+1}\sum_Y\sum_{\l_1\l_2}\frac{|f^{q^{X,i}Y}_{\l\l_1\l_2}|^2}{\hat{C}^{YY}_{\l_1}\hat{C}^{x^ix^i}_{\l_2}}
    \,. \label{Eq:est:norm}
}
The noise spectrum of the reconstructed remote quadrupole corresponds to the estimator normalization, and we use the above equation for computing the noise spectrum. 
The reconstruction noise spectra are computed with a public code of \cite{Philcox:2022:psz}; We first compute the power spectra, $\hat{C}^{x^ix^i}_\l$ and $C^{\delta\tau^ix^i}_\l$, in the Limber approximation since we only use small-scale multipoles for the reconstruction. We assume an LSST-like galaxy survey \cite{LSST} with the same redshift distribution of galaxies, galaxy bias, and the same number density of galaxies as that used in the previous works \cite{Philcox:2022:psz,Lee:2022:pSZ-biref}. We use the multipole between $100$ and $5000$ to compute the noise spectrum. We choose six top-hat redshift bins whose bin widths are equal in comoving distance. 

\begin{figure}[t]
 \centering
 \includegraphics[width=88mm]{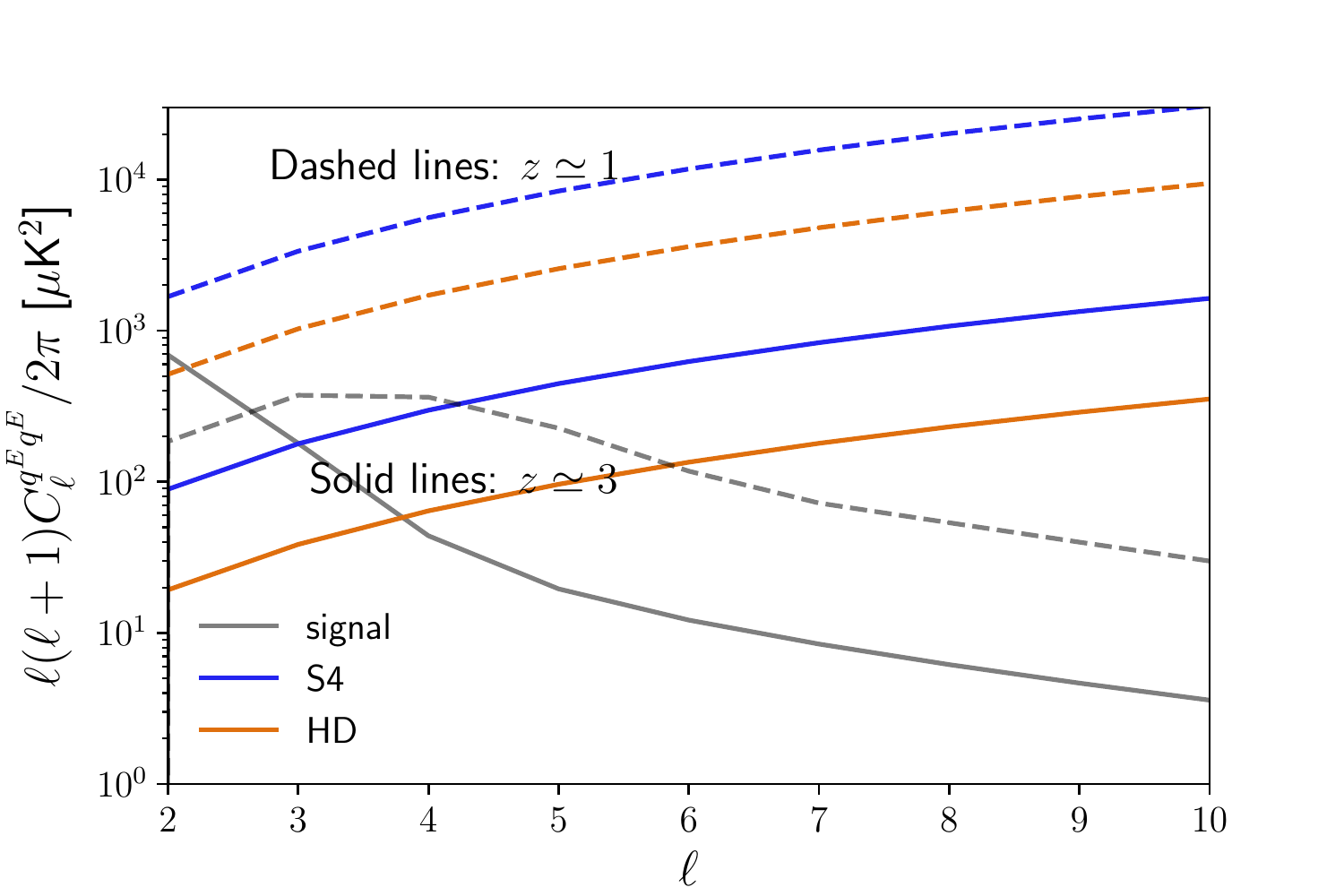}
 \caption{Reconstruction noise power spectrum of the remote-quadrupole $E$-modes, $N_\l^{q^{E,i}q^{E,i}}$, for the third (dashed, $z\simeq1$) and sixth (solid, $z\simeq3$) redshift bin, using high-$\l$ CMB measurements from S4 (blue) / HD (orange) with LSST galaxies. The solid gray lines show the angular power spectra of the remote-quadrupole $E$-modes at each bin.}
 \label{fig:psz-noise}
\end{figure}

Figure \ref{fig:psz-noise} shows the $q^E$ reconstruction noise spectra for the third and sixth bins for S4 and HD cases (see Table \ref{tab:cmbexp} for the experimental setup). Note that the $q^B$ reconstruction noise spectrum is close to that of $q^E$. 
The reconstruction noise power spectrum at the third bin is much larger than the remote quadrupole signals. At the sixth bin, the noise power spectrum is less than the signal power spectrum at $\l \alt 4$. 
We can only use the large-scale remote quadrupole to constrain cosmology.

\subsection{Cosmic birefringence and pSZ} \label{sec:cbtomography}

The cosmic birefringence converts part of the remote quadrupole $E$- to $B$-modes. At a comoving distance, $\chie$, the remote-quadrupole $B$-modes are given in the small-angle limit ($|\beta|\ll 1$) as 
\al{
    q^B_{\l m}(\chi) \simeq 2\beta(\chi)q^E_{\l m}(\chi)
    \,,
}
where the birefringence angle is given by \cite{Carroll:1989:rot,Carroll&Field:1991,Harari:1992} 
\al{
    \beta(\chi) = \frac{g_{\phi\gamma}}{2}[\phi(0)-\phi(\chi)]
    \,. \label{eq: beta0}
}
Here, $\phi(\chi)$ is an ALP field at comoving distance $\chi$. 
This remote quadrupole $B$-modes can be measured by the reconstruction presented in the previous section. The remote-quadrupole $B$-modes also contribute to the total observed CMB $B$-modes: 
\al{
    B^{\rm pSZ}_{\l m} &= -\frac{\sqrt{6}}{10}\INT{}{\chi}{}{0}{\chi_*}g_{\rm vis}(\chi)2\beta(\chi)q^E_{\l m}(\chi)
    \\ 
    &\simeq -\frac{\sqrt{6}}{10}\sum_i2\beta_i\INT{}{\chi}{}{\chi_{i-1}}{\chi_i}g_{\rm vis}(\chi)q^E_{\l m}(\chi)
    \notag \\
    &\equiv \sum_i 2\beta_i E^i_{\l m}
    \,, \label{Eq:BpSZ}
}
where we denote $\beta_i$ as the representative birefringence angle at $i$th bin and introduce the CMB $E$-modes generated during $\chi_{i-1}\leq\chi\leq\chi_i$ as $E^i_{\l m}$. 

Let us derive the auto- and cross-angular power spectra between large-scale CMB $E$-modes ($E_{\l m}$), CMB $B$-modes ($B_{\l m}$), remote-quadrupole $E$-modes ($q^E_{\l m}$), and $B$-modes ($q^B_{\l m}$). In the small angle limit, the CMB $E$-modes and remote-quadrupole $E$-modes are unchanged by the cosmic birefringence. The auto- and cross-angular power spectra between the CMB $E$ and $B$ in the presence of cosmic birefringence are then given by
\al{
    C_\l^{E'E'} &\simeq C^{EE}_\l 
    \,, \label{Eq:EE} \\ 
    C_\l^{E'B'} &\simeq 2\beta_{\rm rei}C^{EE,\rm rei}_\l + \sum_i 2\beta_i C^{EE^i}_\l
    \,, \label{Eq:EB} \\ 
    C_\l^{B'B'} &\simeq \tilde{C}^{BB}_\l
    \,, \label{Eq:BB}
}
where $\beta_{\rm rei}$ is the birefringence angle of polarization sourced at reionization, $C^{EE,\rm rei}_\l$ is the $E$-mode power spectrum generated during reionization, and $\tilde{C}^{BB}_\l$ is the lensing-induced CMB $B$-modes. Since the pSZ signals are significant only at low multipole ($\ell\alt10$), we ignore the recombination signals which are the dominant contributions at high multipole ($\ell\agt 10$). We do not include the lensing effect except in $C_\l^{B'B'}$ since it does not change the power spectra at low-$\l$ \cite{Naokawa:2023}. We use the Python version of {\tt CAMB} \cite{Lewis:1999bs} to compute $C_{\ell}^{EE}$ and $\tilde{C}^{BB}_\l$.

Similarly, the cross-angular power spectra between the CMB polarization and reconstructed remote quadrupole in the presence of cosmic birefringence are then given by
\al{
    C_\l^{E'q^{E',i}} &\simeq C^{Eq^{E,i}}_\l
    \,, \label{Eq:EqE} \\ 
    C_\l^{B'q^{E',i}} &\simeq 2\beta_{\rm rei}C^{Eq^{E,i}}_\l + \sum_j 2\beta_jC^{E^{j}q^{E,i}}_\l
    \,, \label{Eq:BqE} \\ 
    C_\l^{E'q^{B',i}} &\simeq 2\beta_iC^{Eq^{E,i}}_\l 
    \,, \label{Eq:EqB} \\ 
    C_\l^{B'q^{B',i}} &\simeq 0
    \,. \label{Eq:BqB}
}
The remote-quadrupole auto- and cross-angular power spectra are given by
\al{
    C_\l^{q^{E',i}q^{E',j}} &\simeq C^{q^{E,i}q^{E,j}}_\l
    \,, \label{Eq:qEqE} \\ 
    C_\l^{q^{E',i}q^{B',j}} &\simeq 2\beta_j C^{q^{E,i}q^{E,j}}_\l
    \,, \label{Eq:qEqB} \\ 
    C_\l^{q^{B',i}q^{B',j}} &\simeq 0
    \,. \label{Eq:qBqB}
}
Measuring the above power spectra provides information on the ALP field values at each redshift bin. Therefore, the reconstructed remote quadrupole will be a new source for cosmic birefringence tomography.

Note that the second terms of Eqs.~\eqref{Eq:EB} and \eqref{Eq:BqE}, which are responsible for constraining low-$z$ birefringence angles, do not appear in Ref \cite{Lee:2022:pSZ-biref}. This is because they do not divide the contributions to each $z$ bin as in Eq.~\eqref{Eq:BpSZ}. In the next section, we forecast how these terms improve the constraints on the birefringence angles.

\section{Forecast} \label{Sec:forecast}

\begin{figure*}[t]
 \centering
 \includegraphics[width=88mm]{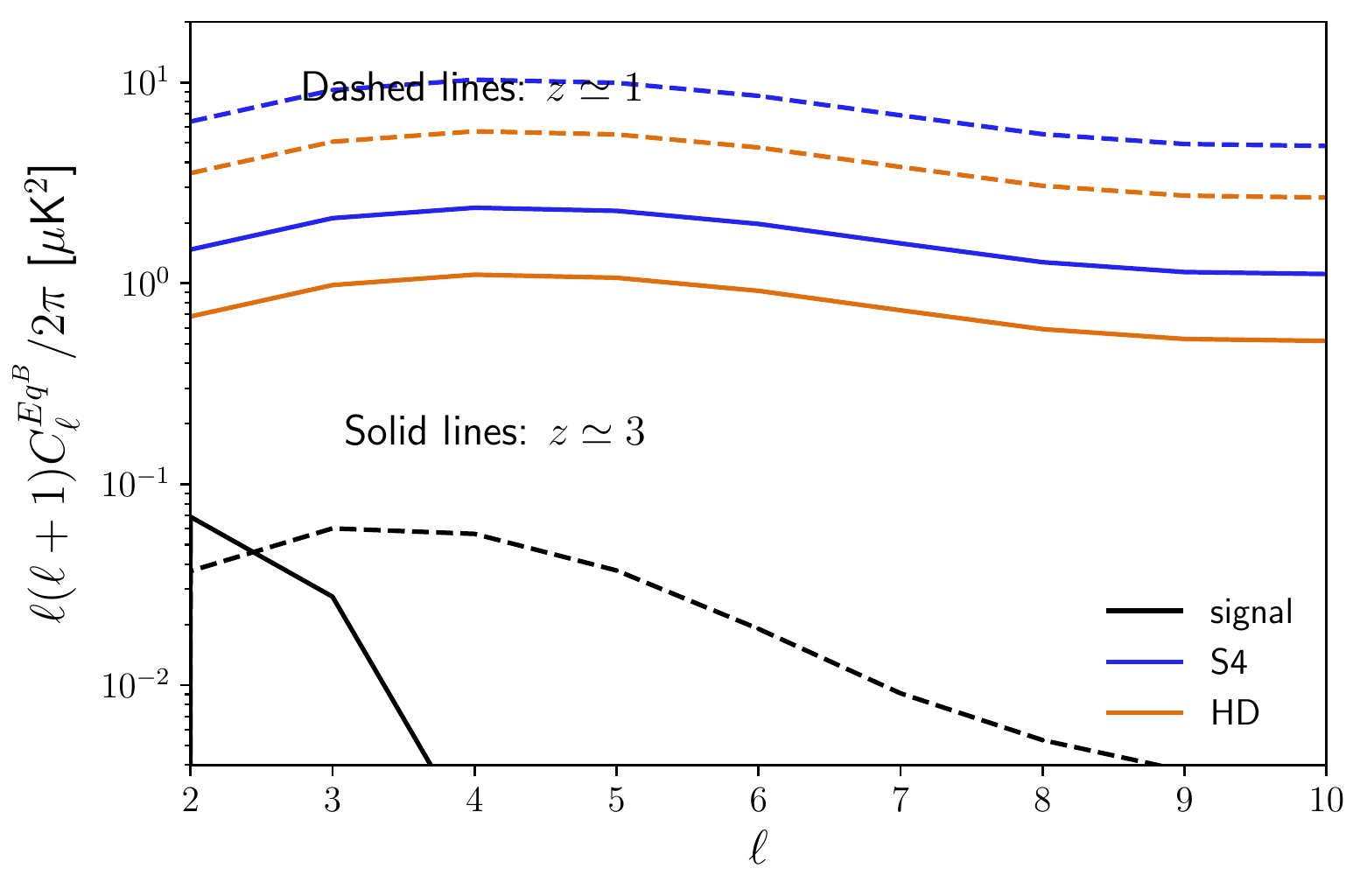}
 \includegraphics[width=88mm]{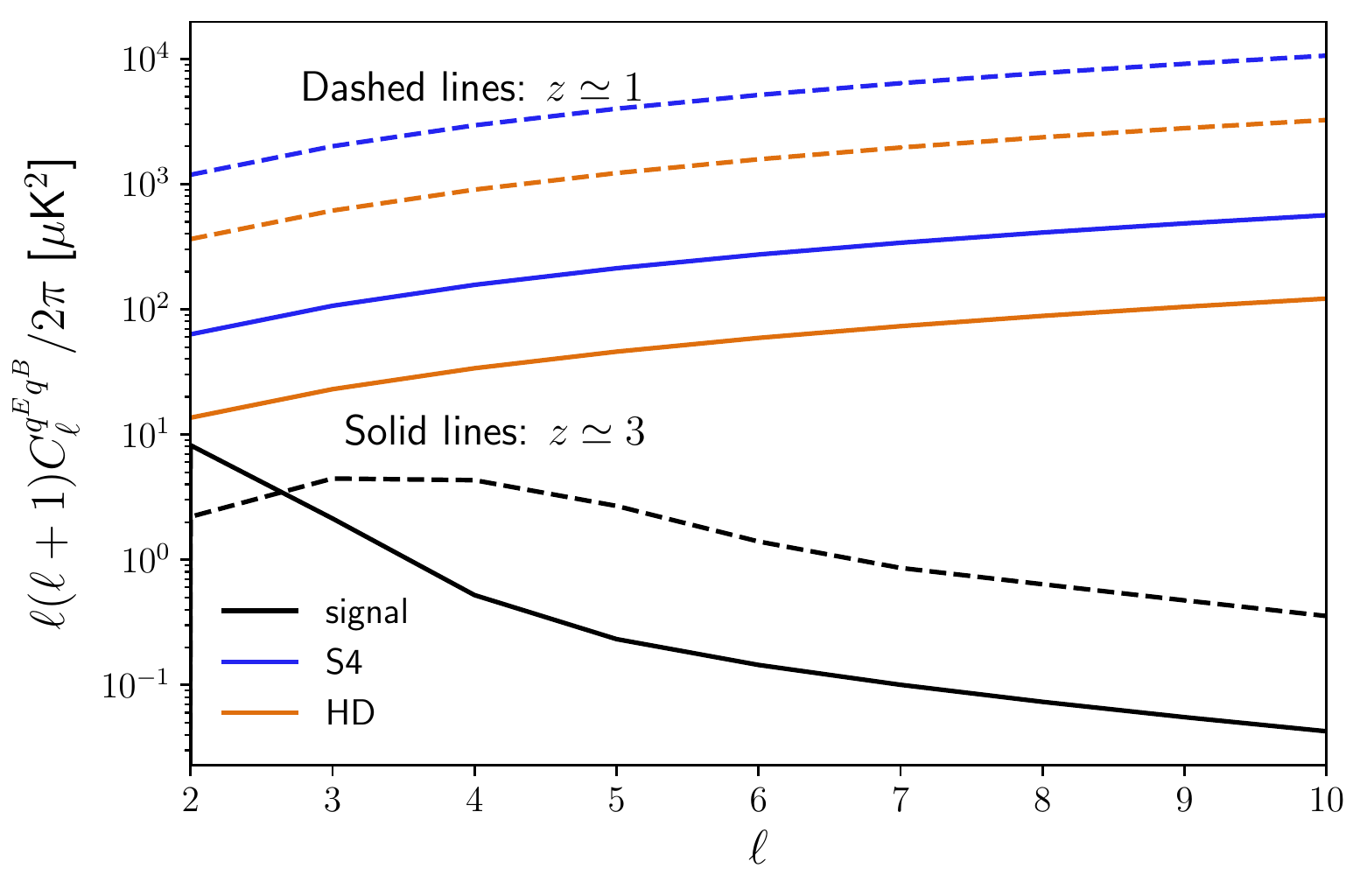}
 \includegraphics[width=88mm]{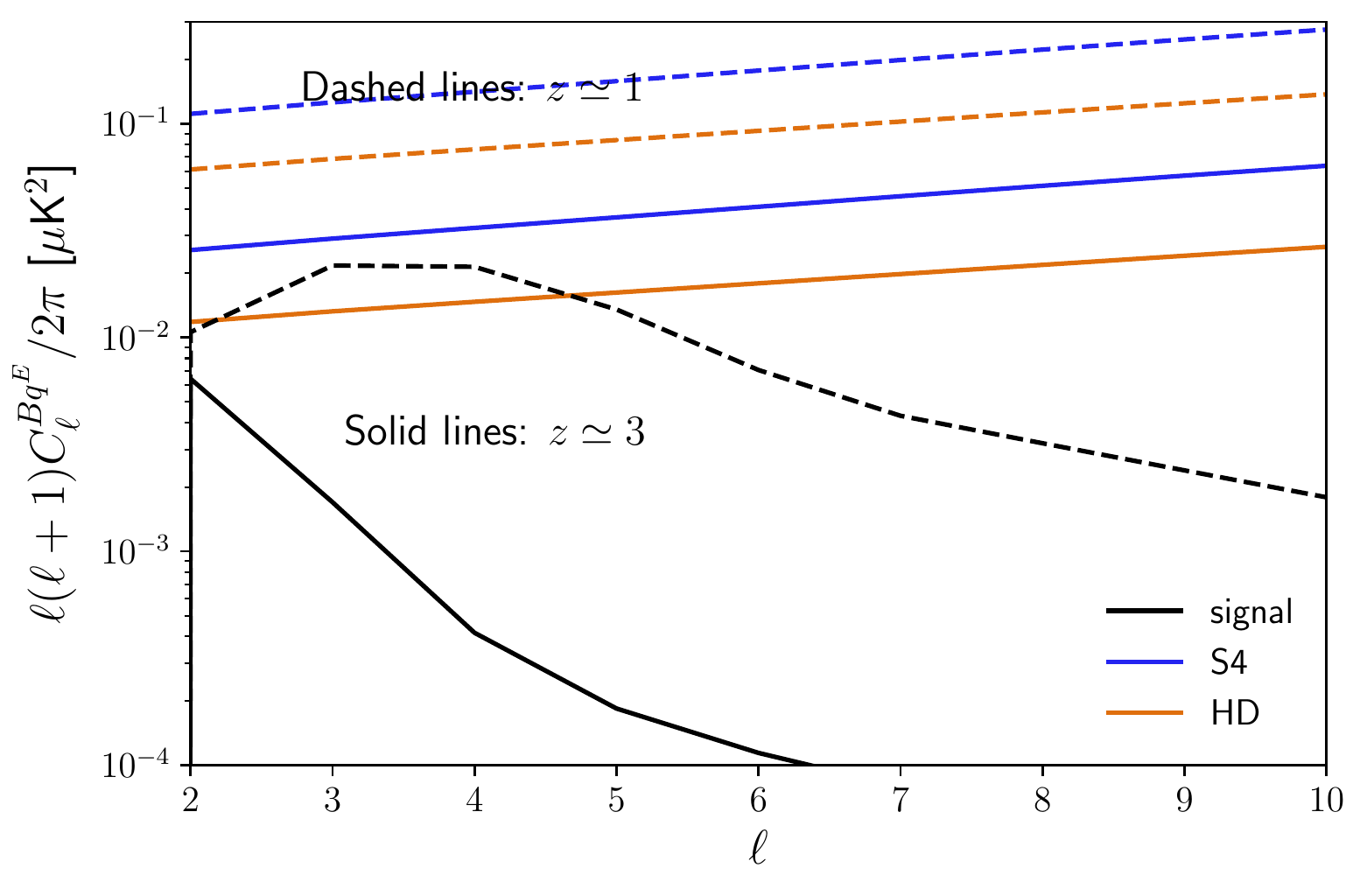}
 \includegraphics[width=88mm]{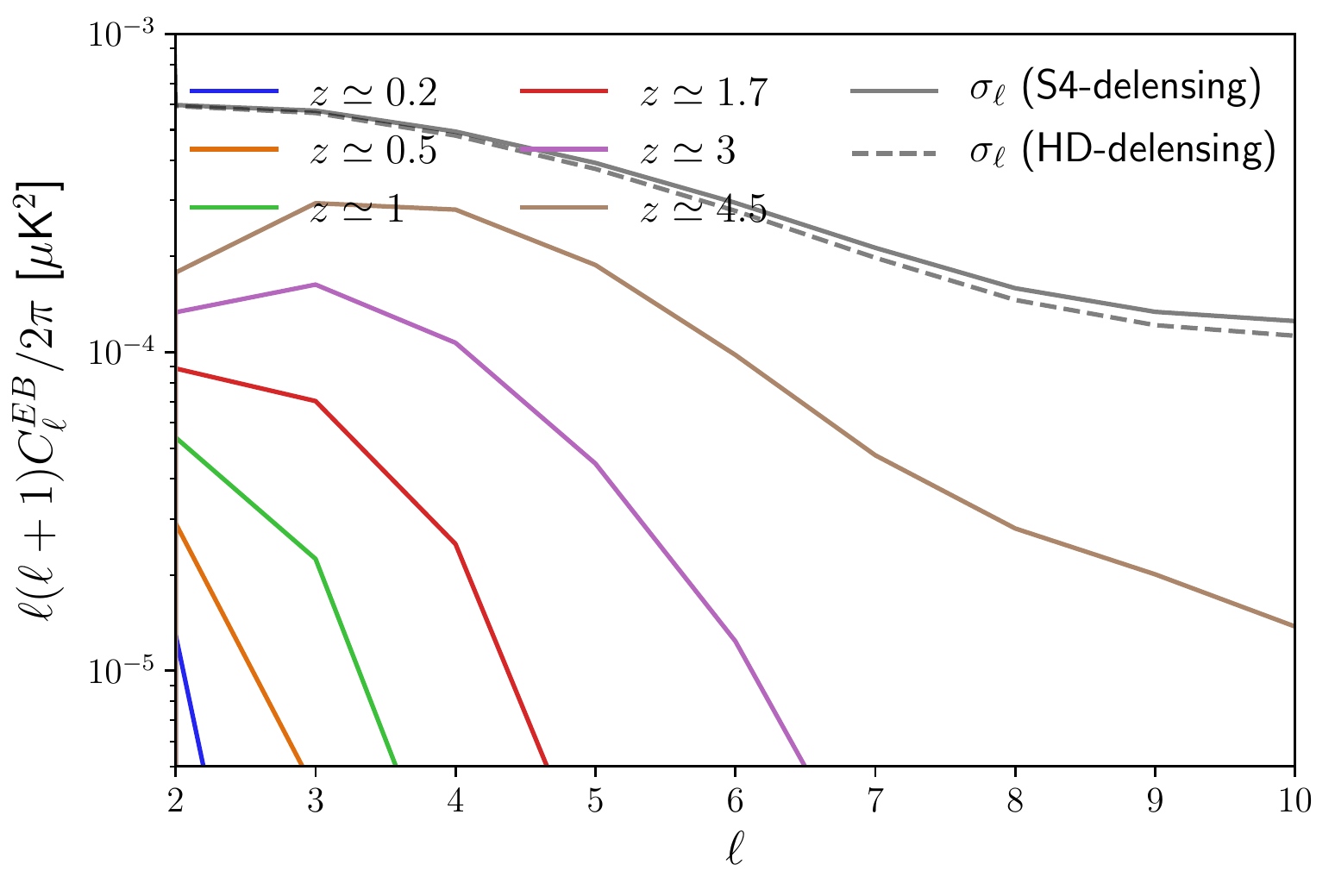}
 \caption{The odd-parity power spectra responsible for constraining rotation angle, $C_\l^{Eq^B}$ (top left), $C_\l^{q^Eq^B}$ (top right), $C_\l^{Bq^E}$ (bottom left), and $C_\l^{EB}$ (bottom right). For $C_\l^{Eq^B}$, $C_\l^{q^Eq^B}$, and $C_\l^{Bq^E}$, we only show the power spectra at the third (solid) and sixth (dashed) redshift bins. For $C_\l^{Eq^B}$, $C_\l^{q^Eq^B}$, and $C_\l^{Bq^E}$, we show the observational errors per multipole defined in Eq.~\eqref{Eq:obserr} for S4 (blue) and HD (orange). For $C_\l^{EB}$, we show two cases of observational errors using S4 (gray solid) or HD (gray dashed) for delensing. The fiducial value of the rotation angle is $0.34$ deg for all spectra. The observational errors are computed with $f_{\rm sky}=0.4$.
 }
 \label{fig:BxqE:ExBpsz}
\end{figure*}

In this section, following \cite{Lee:2022:pSZ-biref}, we estimate the expected constraint on the birefringence angles with the Fisher matrix formalism. We assume that the fiducial values of the birefringence angles are zero. In this case, the small angle limit, $|\beta|\ll 1$ is implicitly assumed for the Fisher matrix formalism. 

We compute the Fisher information matrix as 
\al{
    \{\bR{F}\}_{ij} = \sum_{\l=2}^{\l_{\rm max}} \frac{2\l+1}{2}f_{\rm sky}\Tr\left(\bR{C}^{-1}_\l\PD{\bR{C}_\l}{p_i}\bR{C}_\l^{-1}\PD{\bR{C}_\l}{p_j}\right)\bigg|_{\bm{p}=\bm{p}_{\rm fid}}
    \,. 
}
Here, $\bm{p}$ is a vector containing birefringence angle parameters, $\bm{p}_{\rm fid}$ is the fiducial value, and $f_{\rm sky}$ is the sky coverage of experimental datasets which is set to $0.4$ for our analysis since the wide-field ground-based experiments plan to observe roughly $40\%$ of the sky.
We only need large-angular scales to constrain late-time birefringence and set $\l_{\rm max}=10$. 
$\bR{C}_\l$ is the covariance matrix of observed data and its $(X,Y)$ element is given by
\al{
    \{\bR{C}_\l\}^{XY} = C^{X'Y'}_\l + \delta^{XY} N^{XX}_\l \,,
}
with $X$ and $Y$ are either $E$, $B$, $q^{E,i}$, or $q^{B,i}$. 
We assume that CMB $E$- and $B$-modes are obtained from LiteBIRD, and the remote quadrupole fields are reconstructed by combining S4 or HD with galaxies obtained from LSST. We use the experimental setup for CMB summarized in Table \ref{tab:cmbexp}.

The elements of the signal covariance matrix, $C^{X'Y'}_\l$, are computed from \eq{Eq:EE,Eq:EB,Eq:BB,Eq:EqE,Eq:EqB,Eq:BqE,Eq:BqB,Eq:qEqE,Eq:qEqB,Eq:qBqB}.
Note that, for the lensing $B$-mode spectrum, Eq.~\eqref{Eq:BB}, we multiply a factor $A_{\rm lens}$ to account for the suppression of the lensing $B$-mode by delensing using a lensing map from S4 or HD. 

The noise spectra in the noise covariance, $N^{EE}_\l$ and $N^{BB}_\l$, are computed for LiteBIRD since we only use multipole up to $\l=10$, which is hard to measure from ground-based experiments. In the LiteBIRD noise spectra, we add the residual Galactic foregrounds estimated by \cite{Errard:2016:FG}. The noise spectra of the remote quadrupole field, $N^{q^{E,i}q^{E,i}}_\l$ and $N^{q^{B,i}q^{B,i}}_\l$, are already computed in Sec.~\ref{Sec:pSZ}.

\subsection{Odd-parity power spectra}

The odd-parity power spectra, i.e., $C_\l^{Eq^B}$, $C_\l^{q^Eq^B}$, $C_\l^{Bq^E}$, and $C_\l^{EB}$, constrain the rotation angles of cosmic birefringence. Thus, the high signal-to-noise ratio of these spectra is essential to constrain cosmic birefringence in the late-time universe precisely. 
Figure \ref{fig:BxqE:ExBpsz} shows these odd-parity power spectra with the rotation angle of $0.34$ deg. We also show the observational statistical errors per multipole on each power spectrum, $\sigma_\l$, defined as
\al{
    (\sigma^{XY}_\l)^{-2} \equiv \frac{(2\l+1)f_{\rm sky}(C_\l^{XY})^2}{(C^{XX}_\l+N^{XX}_\l)(C^{YY}_\l+N^{YY}_\l)}
    \,. \label{Eq:obserr}
}
Note that we ignore the cross-power spectrum in the denominator since that contribution is negligible if we assume the rotation angle of $0.34$ deg.
Compared to the odd-parity spectra with $q^B$ (i.e., $C_\l^{Eq^B}$ and $C_\l^{q^Eq^B}$), the cross spectra with the CMB $B$-modes ($C_\l^{Bq^E}$ and $C_\l^{EB}$) have larger signal-to-noise and their measurements provide better constraints on cosmic birefringence at late time. 
For $C_\l^{Bq^{E,i}}$ and $C_\l^{EB}$, the signal power spectra become more significant at higher $z$ bins due to an increase of the electron number density. Compared to the statistical error, high-$z$ birefringence angles are well constrained by $C_\l^{EB}$.

\subsection{Constraints on birefringence angles}

\begin{figure}[t]
 \centering
 \includegraphics[width=85mm]{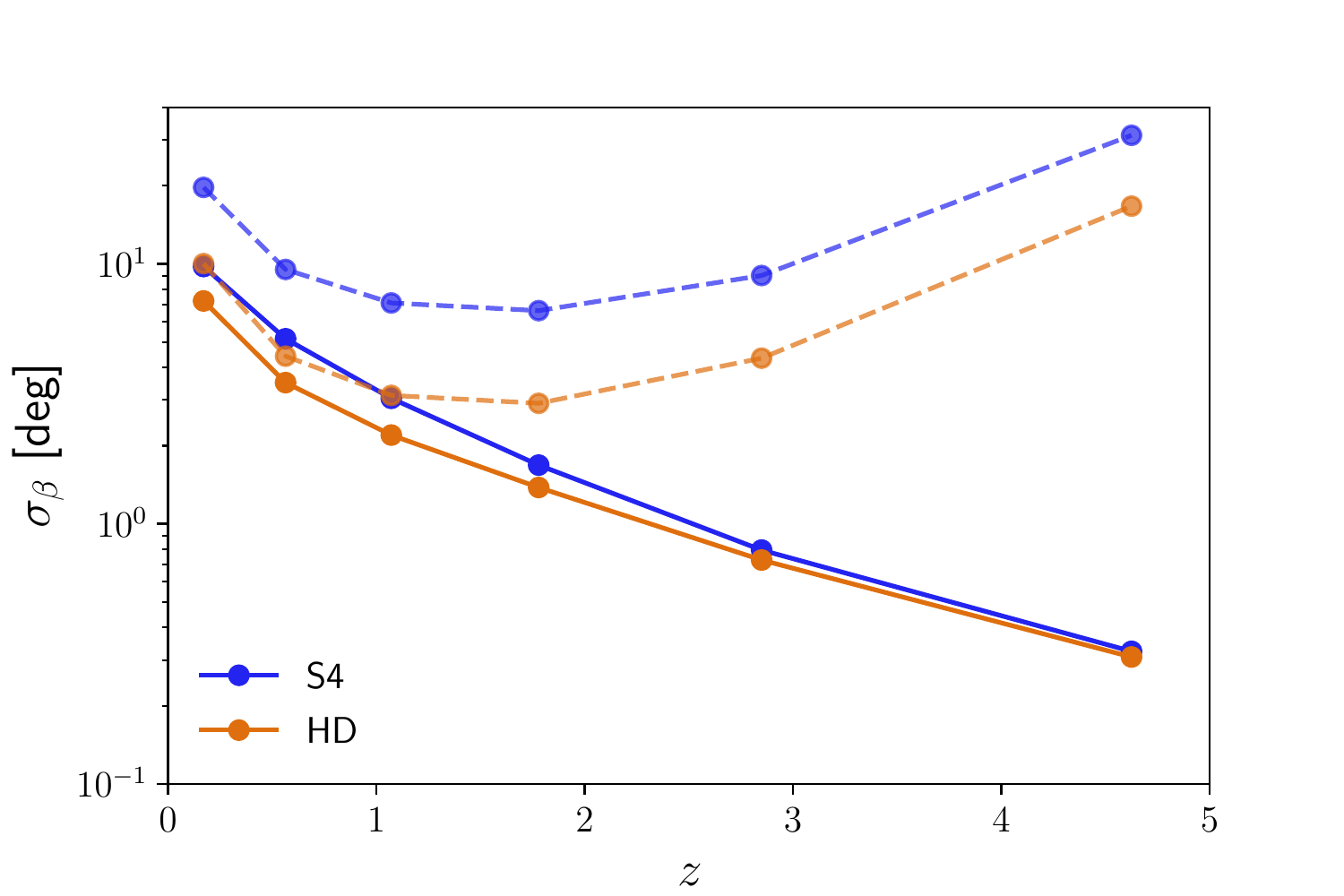}
 \caption{$1\sigma$ constraint on the rotation angles at each redshift bin, including all relevant terms (solid) and ignoring the late-time birefringence effect in the observed $B$-modes (dashed). We assume that the CMB polarization is obtained from a LiteBIRD-like experiment. The remote quadrupole $E$- and $B$-modes are reconstructed from a ground-based S4-like (blue) or HD-like (orange) experiment with an LSST-like galaxy survey.}
 \label{fig:const:pSZ-each}
\end{figure}

We first compute the constraints on the rotation angle at each bin independently. Figure \ref{fig:const:pSZ-each} shows the $1\sigma$ expected constraints on the cosmic birefringence angles at each redshift bin, i.e., $\sigma(\beta_i)\equiv1/\sqrt{\{\bR{F}\}_{ii}}$. We show the cases with S4 and HD for reconstructing the remote quadrupole fields. We also plot the case if we only use part of the odd-parity spectra as in \cite{Lee:2022:pSZ-biref}, i.e., $C_\l^{Eq^{B,i}}$ and $C_\l^{q^{E,i}q^{B,j}}$. The constraints with all the relevant odd-parity power spectra are improved by more than an order of magnitude at high redshift bins compared to the case with only part of the parity-odd power spectra. These results are consistent with the implications obtained from Fig.~\ref{fig:BxqE:ExBpsz}. 

\begin{figure}[t]
 \centering
 \includegraphics[width=85mm]{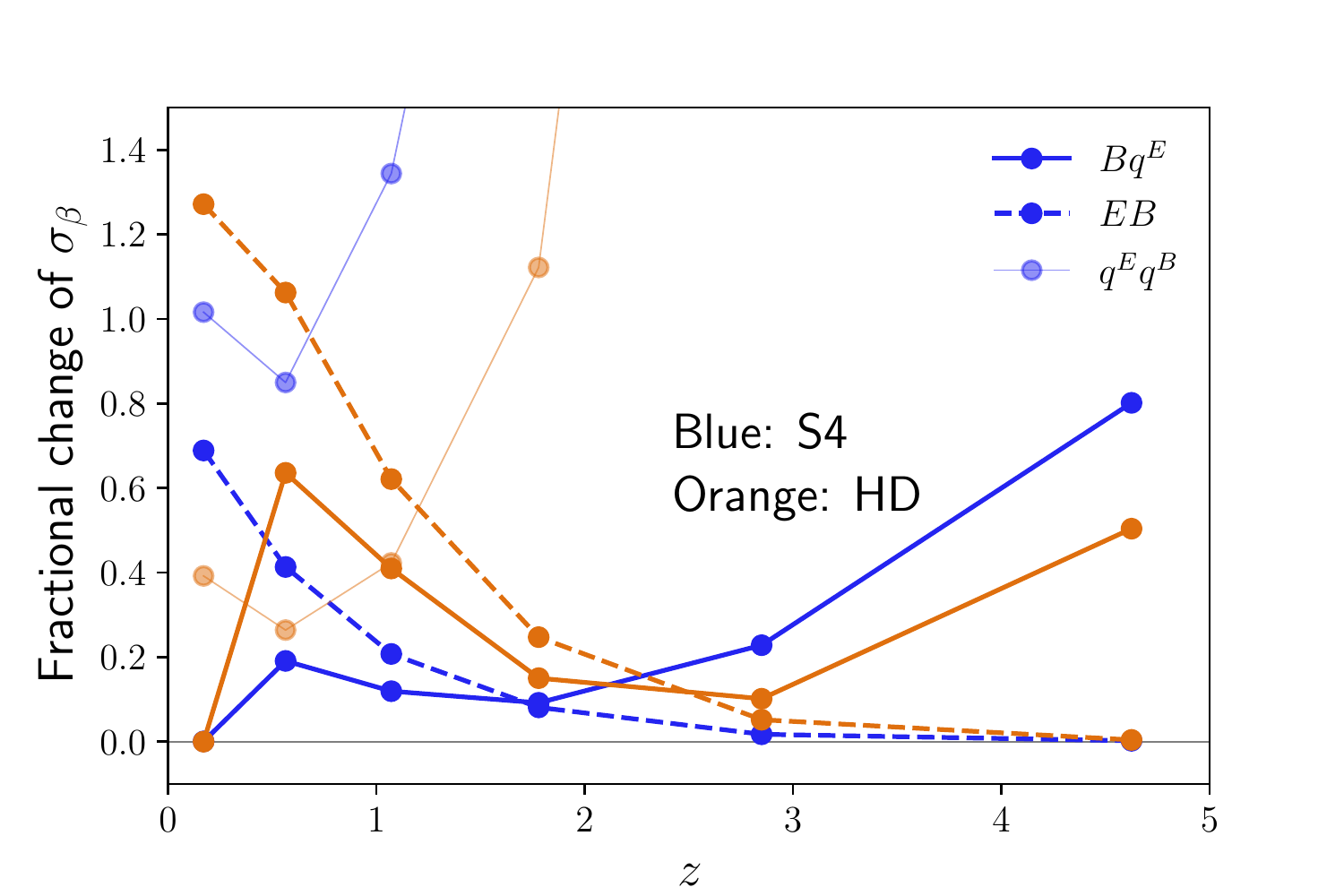}
 \caption{Same as Fig.~\ref{fig:const:pSZ-each} but the fractional change using only large-scale $B$-modes and remote quadrupole $E$-modes ($Bq^E$), large-scale $E$- and $B$-modes ($EB$), and remote quadrupole $E$- and $B$-modes ($q^Eq^B$).}
 \label{fig:const:comp}
\end{figure}

Figure \ref{fig:const:comp} shows the fractional change of $\sigma(\beta_i)$ with only each power spectrum to the case with all power spectra. The case with $C_\l^{Eq^{B,i}}$ is excluded from the figure since the constraint is much worse than in other cases. At high redshift bins, the constraint comes mostly from the $EB$ power spectrum. At lower redshift bins, the $Bq^{E}$ power spectrum dominates the constraint on the birefringence angle. Since the reconstruction noise of the remote quadrupole is much larger than the signal, as shown in Fig.~\ref{fig:psz-noise}, the $q^Eq^B$ cross-power spectrum cannot tightly constrain birefringence angles at any redshifts. However, in the HD case, the remote quadrupole is reconstructed more precisely, and the $q^Eq^B$ power spectrum mildly contributes to constraining the birefringence angles at lower redshift, where other observables also do not tightly constrain the birefringence angles. The constraint from $EB$ power spectrum is $\sim 0.3$ deg at the highest bin. Even if we only use $Bq^E$ power spectrum, the constraint becomes $\sim 0.5$ deg at the highest bin for S4.

\begin{figure}[t]
 \centering
 \includegraphics[width=85mm]{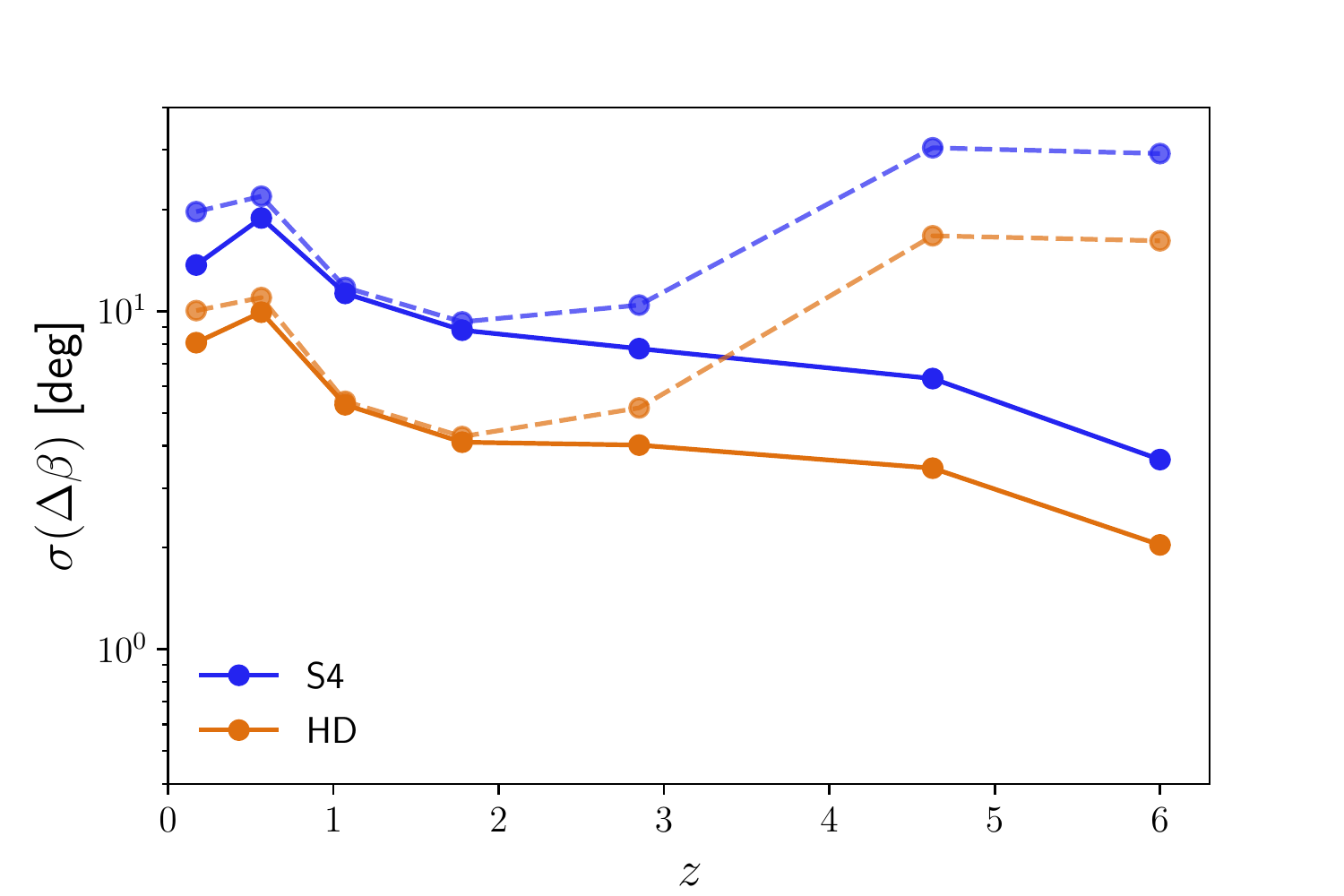}
 \caption{Same as Fig.~\ref{fig:const:pSZ-each} but for $1\sigma$ constraint on the reconstructed rotation angles, $\Delta\beta_i$. We regard the reionization birefringence angle as the rotation angle at $z=6$.}
 \label{fig:const:pSZ-z}
\end{figure}

Next, we show the model-independent joint constraints on the birefringence angles. The parameters, $\beta_i$, are not independent in terms of $\phi$, and following \cite{Lee:2022:pSZ-biref}, we introduce the following parameters:
\al{
    \Delta\beta_i = \beta_i-\beta_{i-1} \,,
}
with $i=2,3,\cdots,n$, $\Delta\beta_1=\beta_1$, and $\beta_n=\beta_{\rm rei}$. The above birefringence angle depends only on the evolution of the ALP fields in each redshift bin. We evaluate the expected $1\,\sigma$ constraint as $\sigma(\Delta\beta_i)\equiv\{\bR{F}^{-1}\}_{ii}^{1/2}$.

Figure \ref{fig:const:pSZ-z} shows the constraint on the reconstructed birefringence angles, $\Delta\beta_i$, for the cases with S4 and HD to reconstruct the remote quadrupole fields. We also show the case if we ignore the contributions of $Bq^E$ and $EB$ cross-power spectra sourced by the pSZ effect. If we use only part of the power spectra, the constraints become very weak at high redshifts.

\subsection{Discussion}

The cosmic birefringence tomography with the pSZ effect is a useful probe of ALP models producing a large birefringence signal in the late-time universe, especially a scenario predicting $|\beta_i| > |\beta_{\rm rec}|$.
While a single-field ALP model does not realize 
such a scenario \cite{Fujita:2020ecn}, this could happen if multiple ALPs exist and each ALP rotates the CMB linear polarization plane, and hence the net birefringence angle we observe is the sum of these angles.

To demonstrate this, we consider the following simple model: two ALP fields $\phi_1$ and $\phi_2$ have periodic potentials generated by the instanton effects
\begin{equation}
m_{\phi_1}^2f_{\phi_1}^2\left[ 1 - \cos\left(\dfrac{\phi_1}{f_{\phi_1}}\right) \right] + m_{\phi_2}^2f_{\phi_2}^2\left[ 1 - \cos\left(\dfrac{\phi_2}{f_{\phi_2}}\right) \right] \ ,
\end{equation}
where $m_{\phi_{1,2}}$ and $f_{\phi_{1,2}}$ are the ALP's mass and decay constant.
Then, introducing the ALP couplings to photon
\begin{equation}
-\dfrac{1}{4}\left(g_{\phi_1\gamma}\phi_1 + g_{\phi_2\gamma}\phi_2\right)F_{\mu\nu}\tilde{F}^{\mu\nu} \ ,
\end{equation}
the total birefringence angle is given by $\beta = \beta_{\phi_1} + \beta_{\phi_2}$.
To find $\beta$, we solve the background dynamics of ALPs.
We take the equation of motion for a homogeneous ALP field as a usual Klein-Gordon equation in cosmology:
\begin{equation}
\ddot{\phi}_{i} + 3H\dot{\phi}_i + V_{\phi_i} = 0 \quad (i=1,2) \ . 
\end{equation}
Regarding the initial field values for ALP fields, we denote them as
\begin{equation}
\phi_{i,\rm ini} = \theta_{\phi_i} f_{\phi_i} \quad (i=1,2)
\end{equation}
with vacuum misalignment angles $\theta_{\phi_{1,2}}$.
The field starts oscillating at a time when the Hubble parameter becomes comparable with ALP mass.
We define $\chi_{i,\rm osc}$ at which $H(\chi_{i,\rm osc}) = m_{\phi_i}$.
For ALP with mass $H_0 \ll m_{\phi_i} \ll H_{\rm rec}$, the current field value $\phi_i(0)$ is much smaller than the value before the oscillation.
Namely, $\beta_{\phi_i}$ is approximately given by
\al{
    \beta_{\phi_i}(\chie \gtrsim \chi_{i,\rm osc}) \simeq -\frac{g_{\phi_i\gamma}}{2}\phi_{i,\rm ini} \quad (i=1,2)
    \,. \label{eq: beta}
}
Then, representing $g_{\phi_i\gamma}$ in terms of \cite{ParticleDataGroup:2022pth}
\begin{equation}
g_{\phi_i\gamma} = \dfrac{\alpha}{2\pi}\dfrac{c_{\phi_i\gamma}}{f_{\phi_i}} \quad (i=1,2) \ ,
\end{equation}
where $\alpha \simeq 1/137$ is QED fine structure constant and $c_{\phi_i\gamma}$ is dimensionless anomaly coefficient, Eq.~\eqref{eq: beta} is reduced to
\al{
    \beta_{\phi_i}(\chie \gtrsim \chi_{i,\rm osc}) \simeq -\frac{\alpha}{4\pi}c_{\phi_i\gamma}\theta_{\phi_i} \quad (i=1,2)
    \,. \label{eq: beta2}
}
Therefore, $\beta$ is determined by the combination of anomaly coefficients and misalignment angles but independent on the decay constants.

For our phenomenological interest, we assume that the axion masses have a hierarchy as $H_0 \ll m_{\phi_1} \ll H_{\rm rei}$ and $H_{\rm rei} \ll m_{\phi_2} \ll H_{\rm rec}$.
At this time, from Eq.~\eqref{eq: beta}, we evaluate $\beta_{\rm rec}$ at the recombination epoch as
\begin{equation}
\beta_{\rm rec} \simeq -\dfrac{\alpha}{4\pi}(c_{\phi_1\gamma}\theta_{\phi_1} + c_{\phi_2\gamma}\theta_{\phi_2}) \ . \label{eq: betarec2}
\end{equation}
We assume that the anomaly coefficients, generically given by the number of charged fermion loops, are of the same order: $c_{\phi_1\gamma} \simeq c_{\phi_2\gamma} = \mathcal{O}(1)$ \cite{Agrawal:2018mkd}.
Hence, if $\theta_{\phi_1}$ and $\theta_{\phi_2}$ are of the same order but have the opposite signs, $|\beta_{\rm rec}|$ becomes small due to the cancellation in Eq.~\eqref{eq: betarec2}.
On the other hand, $\beta_i$ at or after the reionization epoch is approximately given by
\begin{equation}
\beta_{i} \simeq -\dfrac{\alpha}{4\pi}c_{\phi_1\gamma}\theta_{\phi_1} \ , \label{eq: betai}
\end{equation}
where the contribution from $\phi_2$ in Eq.~\eqref{eq: betai} is negligible because it has already decayed away due to the damped oscillation: $m_{\phi_2} \gg H_{\rm rei}$.
Therefore, we could obtain the condition $|\beta_i| > |\beta_{\rm rec}|$ based on this model.
One can also extend this model to an N-field scenario and derive the probability distribution of $|\beta_i| > |\beta_{\rm rec}|$, preferable to the pSZ tomography.
We leave it for future work.

\section{Conclusion} \label{Sec:Conclusion}

We have discussed cosmic birefringence tomography by combining observations of the CMB polarization and remote quadrupole fields. Among the observables we considered, the $EB$ power spectrum most tightly constrains the late-time birefringence angles at high redshifts ($z\agt 2$). The $1\sigma$ constraints from the $Bq^E$ power spectrum are $20\%$ ($80\%$) worse than those from the $EB$ power spectrum at the fifth (sixth) bin. However, the large-scale $EB$ power spectrum might suffer from Galactic foregrounds, and the $Bq^E$ power spectrum provides a useful cross-check for constraining the high-redshift birefringence angles. The remote quadrupole is more sensitive to the low-$z$ birefringence than the $EB$ power spectrum and is a unique probe of the low-$z$ birefringence sources. Precision measurements of the birefringence angles are crucial to get insight into the origin of cosmic birefringence in the late-time universe.


\begin{acknowledgments}
We thank Selim Hotinli, Eiichiro Komatsu, Nanoom Lee, Fumihiro Naokawa, and Hideki Tanimura for their useful comments and discussion. This work is supported in part by JSPS KAKENHI Grant No. JP20H05859 and No. JP22K03682 (T.N.) and No. 19K14702 (I.O.). Part of this work uses the resources of the National Energy Research Scientific Computing Center (NERSC). The Kavli IPMU is supported by World Premier International Research Center Initiative (WPI Initiative), MEXT, Japan.
\end{acknowledgments}

\appendix

\bibliographystyle{mybst}
\bibliography{cite}

\end{document}